\documentstyle[12pt]{article}
\input epsf

\newcommand{\sect}[1]{\setcounter{equation}{0}\section{#1}}

\newcommand\M{{\cal M}}

\newcommand{\ap}{Ann. Phys.}

\newcommand{\cmp}{Comm. Math. Phys.}

\newcommand{\jmp}{J. Math. Phys.}
\newcommand{\np}{Nucl. Phys.}

\newcommand{\pr}{Phys. Rev.}

\newcommand{\rmp}{Rev. Mod. Phys.}

\newcommand{\qed}{$\hfill \Box$}

\begin{document}

\begin{titlepage}

\begin{flushright}
DAMTP-97-105\\
September 1997\\
gr-qc/9709057
\end{flushright}

\vspace{.5in}

\begin{center}
{\LARGE \bf
Topology change from Kaluza-Klein dimensions}\\

\vspace{.4in}
{\large {Radu Ionicioiu}
        \footnote{\it on leave from Institute of Gravitation and
        Space Sciences, 21-25 Mendeleev Street, 70168 Bucharest, Romania}\\
       {\small\it DAMTP, University of Cambridge}\\
       {\small\it Silver Street, Cambridge, CB3 9EW, UK}\\
       {\small\tt email: ri10001@damtp.cam.ac.uk}\\}
\end{center}

\vspace{.5in}
\begin{center}
{\large\bf Abstract}
\end{center}

\begin{center}
\begin{minipage}{4.75in}
{\small
In this letter we show that in a Kaluza-Klein framework we can have arbitrary topology change between the macroscopic (i.e. noncompactified) spacelike 3-hypersurfaces. This is achieved by using the compactified dimensions as a {\it catalyser} for topology change. In the case of odd-dimensional spacetimes (such as the 11-dimensional $\M$-theory) this is always possible. In the even-dimensional case, a sufficient condition is the existence of a closed, odd-dimensional manifold as a factor (such as $S^1, S^3$) in the Kaluza-Klein sector. Since one of the most common manifolds used for compactification is the torus $T^k= S^1 \times \ldots \times S^1$, in this case we can again induce an arbitrary topology change on the 3-hypersurfaces.

\

}
\end{minipage}
\end{center}

\end{titlepage}
\addtocounter{footnote}{-1}

\sect{Introduction}

The study of topology change is important in our search for a quantum theory of gravity. It is widely believed that fluctuations in topology become relevant at the Planck scale. Wheeler \cite{wheeler} was one of the first who promoted the idea that topology should be dynamical at the Planck scale, the now famous {\it foamlike} structure of spacetime. These ideas have been pushed further by Hawking \cite{hawking} who showed that the dominant contribution to the (Euclidean) path integral comes from metrics having one unit of topology per Planck volume.

The purpose of this letter is to propose a mechanism by which we can induce an arbitrary topology change on the (spacelike) 3-hypersurfaces in a Kaluza-Klein theory.

\sect{The results}

In what follows it is important to make the distinction between a topological and a Lorentzian cobordism \cite{yodzis, reinhart}.

A smooth, compact, $n$-dimensional manifold $\M$ is a {\em topological} cobordism if its boundary has 2 disjoint components $\partial \M= \M_0 \sqcup \M_1$, with $\M_0$ and $\M_1$ two smooth, closed, $(n-1)$-dimensional manifolds (possible empty or nonconnected). Two manifolds are topologically cobordant if and only if they have the same Stiefel-Whitney and Pontrjagin numbers (the oriented case) or only the Stiefel-Whitney numbers (the non-oriented case) \cite{wall, milnor}.

A {\em Lorentzian} cobordism is a topological cobordism $(\M;\ \M_0, \M_1)$ together with a nonsingular, unit vector field ${\bf v}$ which is interior normal to $\M_0$ and exterior normal to $\M_1$. In this case we can define a nonsingular Lorentz metric $g^L_{\mu \nu}$ on $\M$

\[ g^L_{\mu \nu} = g^R_{\mu \nu} - 2 v_\alpha v_\beta \]
where $g^R_{\mu \nu}$ is a Riemann metric on $\M$. This is always possible, since there is a one-to-one correspondence between nonsingular vector fields ${\bf v}$ and Lorentz metrics $g^L_{\mu \nu}$ on $\M$. With respect to $g^L_{\mu \nu}$, $\M_0$ and $\M_1$ are spacelike and we will denote them as the initial and final hypersurfaces of the cobordism.

The central point in any study of topology change follows from a theorem of Geroch \cite{geroch}, which proved that in the absence of metric singularities or closed timelike curves, the spacetime manifold is topologically trivial $\M_0 \times [0,1]$, with $\M_0$ the initial spacelike hypersurface.
The proof relies on the fact that the integral curves of the timelike vector field ${\bf v}$ determine a diffeomorphism between the initial and the final (spacelike) hypersurface, $\M_0 \cong \M_1$ and hence there is no topology change.

Therefore, topology change implies either {\em closed timelike curves} (CTCs for short) or {\em singularities} in the metric. At this point we are faced with a choice between these two options. For the rest of this article we assume that the metric is nonsingular, and hence we are forced to admit the existence of CTCs. The other alternative, in which we have singular metrics but no CTCs, will be studied elsewhere \cite{ri}.

We might ask how serious is such a causality violation, if we admit CTCs in the theory? There are a number of solutions of Einstein equations which {\em do} have CTCs -- the G\"odel solution \cite{godel}, and, in general, many of the vacuum rotating solutions \cite{hawel}.
We do not view this as an argument in favour of CTCs, but since a discussion on their existence is beyond the purpose of the present letter, we will simply accept this as a working hypothesis.

We next require the following theorem of Sorkin \cite{sorkin}.

\

{\bf Theorem 1} (Sorkin, 1986)

Suppose $\M$ is a topological cobordism between $\M_0$ and $\M_1$, $dim \M=n$ and $\M$ is connected. Then $\M$ is a Lorentzian cobordism if and only if

\begin{eqnarray}
\chi(\M)=0\ , \ \ n\ {\rm \ even} \label{s1} \\
\chi(\M_0)=\chi(\M_1)\ , \ \ n\ {\rm \ odd} \label{s2}
\end{eqnarray}

\

We would like to apply this to the case of a Kaluza-Klein theory. The idea is to mediate a topology change on a (spacelike) 3-manifold by using the extra dimensions as a {\it catalyser}. 
The main result is the following theorem.

\

{\bf Theorem 2}

Consider a Kaluza-Klein theory with the spacetime manifold

\begin{equation}
\M^{n+4}=\M^4 \times \M^n
\end{equation}
where $\M^4$ is the usual (macroscopic) spacetime and $\M^n$ is the compactified Kaluza-Klein manifold.
Then $\M^{n+4}$ is a {\it Lorentzian} cobordism between the initial \mbox{$\M^3_0 \times \M^n$} and the final \mbox{$\M^3_1 \times \M^n$}, with $\M^3_0\ , \M^3_1$ arbitrary 3-folds, if

(a) $n=2k+1$\\
or

(b) $n=2k$ and $\chi(\M^n)=0$.

\

{\em Proof:}\\
(a) If the dimension is odd, $n=2k+1$, then in order to have a Lorentzian cobordism we need, from (\ref{s2})

\[ \chi(\M^3_0 \times \M^{2k+1}) = \chi(\M^3_1 \times \M^{2k+1}) \]
but this is trivially satisfied, from the factorization propertiy of the Euler characteristic $\chi(A \times B)=\chi(A) \chi(B)$ and from $\chi(\M^{2k+1})=0$ (closed, odd-dimensional manifold). Then $\M^{n+4}$ is a Lorentzian cobordism between two spacelike hypersurfaces $\M_0= \M^3_0 \times \M^n$ and $\M_1= \M^3_1 \times \M^n$. Since $\M^3_0$ and $\M^3_1$ can be any 3-manifolds, this means that we can have an arbitrary change of topology in the macroscopic 3-space sector.

\

(b) If the dimension is even, $n=2k$, from (\ref{s1}) we need

\[ 0=\chi(\M^{n+4})= \chi(\M^4) \chi(\M^n) \]
and therefore the necessary and sufficient condition for an arbitrary topology change in the macroscopic $\M^4$ sector is $\chi(\M^n)=0$.
A sufficient condition for this to happen is that $\M^n$ should have a factor which is a closed odd-dimensional manifold (such as $S^1$, $S^3$ or an arbitrary torus $T^k$). As this is usually the case (a torus is one of the most common manifolds used for compactification), this means that in such a theory an arbitrary topology change can be induced between the macroscopic 3-spaces. \qed

\

{\em Observation:} The cobordism between $\M^3_0 \times \M^n$ and $\M^3_1 \times \M^n$ always exists, since $\M^3_0$ and $\M^3_1$ are cobordant, being 3 manifolds.

\

Consider a few examples.\\
1) In the simplest, Abelian, Kaluza-Klein theory, the spacetime is

\begin{equation}
\M^5=\M^4 \times S^1
\end{equation}
where $\M^4$ is the usual 4-dimensional spacetime.
The spacelike sections are $\M^3 \times S^1$, with $\M^3$ an arbitrary closed 3-manifold. $\M^5$ is a Lorentzian cobordism, since 

\[ \chi(\M^3_0 \times S^1)= \chi(\M^3_1 \times S^1)= 0 \]
for any 3-manifolds $\M^3_0$ and $\M^3_1$.

The same result holds in any 11-dimensional theory ($\M$-theory, 11D supergravity etc), since the spacelike hypersurfaces are again product manifolds $\M^3\times \M^7$ and $\chi(\M^7)=0$ for any closed $\M^7$. Thus $\chi(\M_0)=\chi(\M_1)$ as needed.

2) In the case of a string theory in $n=10$ dimensions, in order to induce the required topology change on the 3-space, we need to have $\chi(\M^6)=0$. This is not automatically satisfied for any $\M^6$, so in the general case we do not have topology change. However, since usually the compactification of string theory is done on tori $T^k$ or other manifolds having at least one $S^1$ factor, in this case the conditions of {\em Theorem 2} are satisfied and the result is topology change on the 3-space.

\sect{Conclusions}

In this letter we proposed a mechanism for topology change which uses the compactified dimensions in a Kaluza-Klein theory as {\it catalysers} for topology change in the macroscopic, noncompactified sector.

Since any odd-dimensional spacetime can induce this, an obvious candidate is $\M$-theory, which is 11-dimensional. Therefore, in this framework, an arbitrary topology change can occur in the macroscopic sector, irrespective of the Kaluza-Klein manifold used for compactification.

In the case of a 10-dimensional string theory, the condition to have the desired topology change in the macroscopic sector is $\chi(\M^n)=0$. This can be achieved if $\M^n$ contains as a factor a (closed) odd-dimensional manifold (such as $S^1$ or $S^3$). Since the torus $T^k=S^1 \times \ldots \times S^1$ is one of the most used manifolds for compactification, in this case we can induce again the (arbitrary) topology change on the usual 3-space.

We hope that the proposed mechanism of Kaluza-Klein catalysis can be extended in the future from topology change to other processes as well.

\

{\Large\bf Acknowledgements}

\

I am grateful to Dr~Ruth Williams and Dr~Dennis Barden for helpful discussions and support. This work has been kindly supported by Cambridge Overseas Trust, the Ra\c tiu Foundation and ORS.

\end{document}